\documentclass{article}
\usepackage[cmex10]{amsmath}
\usepackage{spconf, amssymb, graphicx}
\usepackage{paralist, url}
\usepackage{algpseudocode}
\usepackage{algorithm}
\usepackage{epsfig, bm}
\usepackage{xcolor}
\usepackage{multirow}

\usepackage[symbol]{footmisc}

\usepackage{newtxtext,newtxmath}
\usepackage{times}

\makeatletter
\def\thickhline{%
  \noalign{\ifnum0=`}\fi\hrule \@height \thickarrayrulewidth \futurelet
   \reserved@a\@xthickhline}
\def\@xthickhline{\ifx\reserved@a\thickhline
               \vskip\doublerulesep
               \vskip-\thickarrayrulewidth
             \fi
      \ifnum0=`{\fi}}
\makeatother

\newlength{\thickarrayrulewidth}
\def\past{{\rm L}}
\def\current{{\rm C}}
\def\future{{\rm R}}

\ninept

\title{Continuous speech separation: dataset and analysis}
\name{Zhuo Chen$^{\ast}$, Takuya Yoshioka$^{\ast}$\thanks{$^\ast$ Equal contributions.}, Liang Lu, Tianyan Zhou, Zhong Meng, Yi Luo$^{\dagger}$,  }
\secondlinename{Jian Wu$^\dagger$\thanks{$^\dagger$ Work performed during internship at Microsoft.}, Xiong Xiao, Jinyu Li}
\address{Microsoft, One Microsoft Way, Redmond, WA, USA}

\begin{document}
\maketitle

\begin{abstract}
This paper describes a dataset and protocols for evaluating continuous speech separation algorithms. Most prior speech separation studies use pre-segmented audio signals, which are typically generated by  mixing speech utterances on computers so that they \emph{fully} overlap. Also, the separation algorithms have often been evaluated based on signal-based metrics such as signal-to-distortion ratio. However, in natural conversations, speech signals are continuous and contain both overlapped and overlap-free regions. In addition, the signal-based metrics only have weak correlation with automatic speech recognition (ASR) accuracy. Not only does this make it hard to assess the practical relevance of the tested algorithms, it also hinders researchers from developing systems that can be readily applied to real scenarios. In this paper, we define continuous speech separation (CSS) as a task of generating a set of non-overlapped speech signals from a \textit{continuous} audio stream that contains multiple utterances that are \emph{partially} overlapped by a varying degree. A new real recording dataset, called LibriCSS, is derived from LibriSpeech by concatenating the corpus utterances to simulate conversations and capturing the audio replays with far-field microphones. A Kaldi-based ASR evaluation protocol is established by using a well-trained multi-conditional acoustic model. A recently proposed speaker-independent CSS algorithm is investigated by using LibriCSS. The dataset and evaluation scripts are made available to facilitate the research in this direction\footnote{https://github.com/chenzhuo1011/libri\_css}.
\end{abstract}
\begin{keywords}
Continuous speech separation, automatic speech recognition, LibriCSS, overlapped speech, permutation invariant training
\end{keywords}

\section{Introduction}
\label{sec:intro}
\vspace{-.5em}

As a natural phenomenon in human interactions, overlapping speech occupies a significant part of conversation time. This poses challenges for many speech technologies including automatic speech recognition (ASR) and  speaker diarization because they usually assume one or zero speaker to be active at the same time. Speech separation could provide a solution for this problem. 

The speech separation technology has been significantly improved over the past five years
by leveraging deep learning. 
One fundamental challenge in overlapped speech separation is the inherent indeterminacy of the speaker order, i.e. the permutation problem~\cite{hershey2016deep}, which complicates supervised model training.  
To deal with this problem, 
\cite{hershey2016deep} proposed deep clustering (DC), achieving high quality single-channel speech separation for the first time by using  a recurrent network with an affinity-based objective function that is invariant to the number of speakers and their permutation. In \cite{yu2017permutation}, the authors proposed permutation invariant training (PIT), which was shown to achieve a similar level of  separation performance by exhaustively searching for the best speaker permutation during model training. Numerous extensions to these methods were proposed with different focuses~\cite{wang2018alternative,xu2019optimization,chen2017deep,chen2018sequence,li2019listening,wang2019pitch}.

Another approach to the permutation indeterminacy problem is ``informed extraction'', which makes  use of additional information to distinguish a target speaker from other participanting speakers. 
Use of visual information~\cite{ephrat2018looking,zhao2018sound}, audio snippets of the speakers~\cite{zmolikova2017speaker,wang2018voicefilter,xiao2019single}, or their locations~\cite{chen2018location,perotin2018multichannel,zhao2018two} were investigated. 
On top of, or aside from, these algorithmic improvements, 
researchers also sought more effective input features~\cite{wang2018multi,yoshioka2018multi} and model architectures~\cite{luo2018tasnet,luo2019conv,shi2019furcanext}. 

The signal-to-distortion ratio (SDR) \cite{fevotte2005bss_eval} and the scale-invariant signal-to-noise ratio (SISNR) \cite{le2019sdr} have been steadily increasing on WSJ0-2mix, the most widely used speech separation dataset, which indicates the consistent progress of the separation technology. An early system \cite{hershey2016deep} achieved SDR improvement of 6.3 dB
while \cite{luo2019dualpath} improved the SDR by 19.0 dB.
\cite{luo2019conv} reported that, in WSJ0-2mix, separated speech signals generated by TasNet, a state-of-the-art separation method, were almost indistinguishable from clean utterances.

Despite those advances, existing speech separation evaluation schemes have several shortcomings that make it difficult to assess the practical relevance of the tested algorithms. Firstly, while most separation studies focus on disentangling fully overlapped speech signals, it is crucial that separation algorithms do not introduce signal distortion when one person is talking. An overlap ratio is usually below 20\% in a natural meeting \cite{Cetin06}. Therefore, the evaluation of separation algorithms has to be done in a way that considers both the separation accuracy for the overlapping periods and the distortionlessness for the overlap-free segments, which was not considered sufficiently before.

This also leads to the second issue. Most existing speech separation evaluation schemes use pre-segmented samples by implicitly assuming an accurate overlap detector to be available. Some methods further require prior knowledge of the number of participating speakers or knowledge of a target speaker. 
However, in practice, obtaining such information from conversational recordings is challenging. 
In addition, some of these elements must be interconnected by nature. For example, accurate speaker counting would benefit from separating each person from the mixture. Focusing only on the separation accuracy could make us blind to this correlation between different elements. 

In addition, most speech separation methods have been evaluated in terms of signal-based metrics such as SDR or SISNR. However, it is known that the signal-level performance metrics are only weakly correlated with the ASR accuracy or the perceptual sound quality. 

To bring the speech separation technology to a real world task, 
\cite{yoshioka2018multi} proposed continuous speech separation (CSS), i.e., generating multiple overlap-free signals from an input audio stream that occasionally contains overlapped utterances. 
In this paper, we create a new dataset, called LibriCSS, recorded by using LibriSpeech utterances to facilitate CSS research while attempting to keep the task simple. Based on this dataset, we explore different aspects of the CSS method of \cite{yoshioka2018multi}. The dataset is available at https://github.com/chenzhuo1011/libri\_css along with Kaldi-based ASR evaluation scripts.



\section{Continuous speech separation}
\label{sec:css}
\vspace{-.5em}

In the most general way, continuous speech separation (CSS) can be defined as a process of generating a set of overlap-free speech signals from a \textit{continuous} audio stream consisting of multiple utterances spoken by different people, which can be hours long and contain both overlapped and overlap-free parts.
In conversation scenarios like meetings, adjacent utterances overlap only by 10--15\%~\cite{Cetin06} on average. 
When used as a front-end module, CSS allows downstream speech applications, such as ASR or speaker diarization, to operate on the assumption that only one person is active at each time point. 
Note that this task implicitly includes overlap detection as a subtask.

There are multiple approaches to CSS. 
One approach could be to keep generating an enhanced signal in a continuous fashion for each involved speaker as in \cite{HoriMeeting,NTTAllNeural}, which requires online speaker diarization. Alternatively, this could be achieved with offline processing by first performing speaker diarization and then extracting individual speaker signals by using an informed extraction approach based on the speaker embedding vectors~\cite{Kanda2019}. 
The detect-then-separate strategy, i.e., performing overlap detection prior to separation, is an approach that many previous separation studies implicitly assumed and that requires prior knowledge of the boundaries of each utterance.
One potential problem with this approach is that separated signals for the overlapped segments need to be concatenated with the preceding and following signals in a way that keeps speaker consistency, which is not a trivial task. 
Also, overlap detection and speech separation are essentially inter-dependent problems. Thus, the sequential approach would lead to a sub-optimal solution.

In this paper, we investigate the speaker-independent CSS approach proposed in \cite{yoshioka2018multi}.
With this approach, we output a fixed number ($N$) of audio streams, where each stream contains at most one active speaker at any time. 
For segments with no speaker overlaps, this CSS algorithm routes the incoming speech into one of the output channels,  while the other output channels produce zero or negligible noise. The method was applied to real meeting recordings, where $N$ was set to two because three-fold overlaps rarely happen in meetings. 
It yielded significant ASR accuracy gains for real meetings compared with conventional beamformers~\cite{PrincetonASRU2019}.

\section{LibriCSS}
\label{sec: data}
\vspace{-.5em}

\begin{figure}
\centering
\includegraphics[scale=0.4]{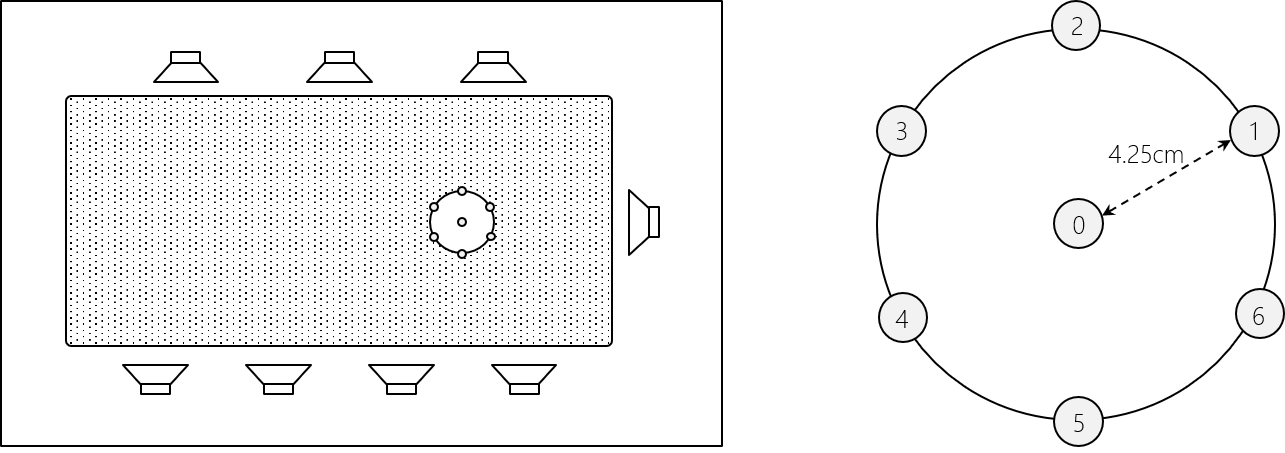}
\vspace{-1em}
\caption{Example recording setup (left) and microphone array geometry (right).}
\label{fig: data}
\end{figure}





\subsection{Dataset}
\vspace{-.3em}

The LibriCSS dataset is aimed at facilitating speech separation algorithm evaluation in the continuous input setting, thereby bridging the gap between the state of the speech separation research and what is required in real world applications. At the same time, the dataset is designed to be simple enough to be broadly accessible.
It consists of multi-channel audio recordings of ``simulated conversations'', each containing multiple utterances spoken by different speakers. Each utterance is taken from LibriSpeech and played back from a loudspeaker placed in a room. We refer to each simulated conversation as a session in the following. 

The dataset is designed to capture three features that are mostly missing in existing popular corpora for speech separation.
Firstly, the data are recorded in a room instead of being generated by simulation. 
The simulated data tend to oversimplify room acoustics especially in multi-channel scenarios. 
Secondly, the dataset encompasses different overlap ratios and silence settings to help analyze how different algorithms work under various overlap conditions. As discussed in Section \ref{sec:intro}, many existing studies only consider separating sufficiently overlapped utterances, overlooking the possibility of introducing signal distortion in non-overlapped segments. 
Thirdly, the audio signals are continuously recorded to enable CSS evaluation. 
Meanwhile, the ground-truth segmentation is also provided, allowing for the conventional utterance-wise evaluation.

LibriCSS consists of 10 hours of audio recordings.
10 sessions are included in the dataset, where each session is approximately one hour long. Each session is made up of six 10-minute-long ``mini sessions'' that have different overlap ratios (OVRs), ranging from 0 to 40\%, where $\text{OVR}= L_{\text{ovl}} / L_{\text{all}}$, with $L_{\text{ovl}}$ and $L_{\text{all}}$ being the total overlapped region length and the total speech length, respectively. 
Each mini session includes eight speakers that are randomly selected from 40 speakers in the LibriSpeech ``test clean'' set. The total number of utterances in each mini session ranges from 52 to 125.
For systems that do not perform speaker diarization before ASR, quick turn-taking is likely to result in concatenating multiple speaker utterances. This could end up with carrying over the internal state of a preceding speaker to the next speaker during decoding processing, which might result in ASR performance degradation.
To investigate this, for the 0\% overlap scenario, two conditions are considered with respect to the silence length between utterances. In both conditions, the utterances are played back sequentially without overlaps. In the short silence version, the inter-utterance silence length is randomly sampled from 0.1 to 0.5 seconds. The long silence version uses an inter-utterance silence period of 2.9--3.0 seconds. 

The recordings were made in a regular meeting room 
by using a seven-channel circular microphone array (see Fig.~\ref{fig: data}). 
The loudspeaker locations were randomly chosen in the meeting room while they remained the same for each session. The loudspeaker-to-microphone distances ranged from 33 cm to 409 cm. 

\subsection{Related Work}
\vspace{-.3em}

With the awareness of the limitations of the existing speech separation evaluation schemes, several datasets were recently used other than WSJ0-2mix, focusing on different aspects of the speech separation problem.
In \cite{wichern2019wham}, the authors collected replayed WSJ0 utterances in different outdoor environments to evaluate the noise robustness of speech separation systems. ASR evaluation was not considered. 
\cite{bahmaninezhad2019comprehensive} used artificially simulated multi-channel WSJ0 utterances. While the paper reported word error rates (WERs), it used an extremely simple ASR system trained on clean utterances, making it hard to assess the usefulness of the separation algorithms. 
In addition, both datasets were collected and evaluated in an utterance-wise manner, with oracle utterance segmentation.

There are several corpora consisting of conversational recordings including natural speech overlaps. 
In \cite{barker2018fifth,van2019dipco}, ``dinner party'' style data were collected. 
However, systems were allowed to make use of oracle speaker segmentations. The evaluation was also performed in an utterance-wise fashion. 
Meetings are other situations where overlaps naturally happen~\cite{ICSICorpus,AMICorpus}. 
While the meeting corpora can be used for evaluating the speech separation algorithms in an end-to-end fashion as in \cite{PrincetonASRU2019}, it is desirable to have a dataset that focuses on speech separation evaluation and allows for detailed analysis. 

Our LibriCSS dataset were collected under the same setup as \cite{PrincetonASRU2019}. 
The full training and ASR evaluation setup will be provided. 
This will allow speech separation algorithms to be easily tested in a more practical setting than \cite{wichern2019wham,bahmaninezhad2019comprehensive} while keeping the evaluation scheme simple.

\subsection{Evaluation protocol}
\vspace{-.3em}

\subsubsection{ASR setup}
\vspace{-.5em}

We use an ASR system to measure the speech separation accuracy. Our acoustic model is trained on 960 hours of the LibriSpeech training data, which contains both clean and noisy audio. Kaldi~\cite{povey2011kaldi} is used to generate a phonetic decision tree and alignments. A bidirectional long short term memory (BLSTM) acoustic model is built using PyKaldi2~\cite{lu2019pykaldi2}, a toolbox developed on top of Kaldi and PyTorch. The BLSTM has 3 layers, and each layer has 512 cells for both forward and backward directions. The model is initialized with cross-entropy training, followed by fine-tuning with maximum mutual information (MMI) training. Decoding is performed with the standard 4-gram langauge model for LibriSpeech. Refer to~\cite{lu2019pykaldi2} for more details about the ASR setup.

Two evaluation configurations are considered: one uses pre-segmented audio and one uses unsegmented audio. These are referred to as utterance-wise evaluation and continuous input evaluation, respectively. The latter is used for evaluating CSS algorithms. 

\subsubsection{Utterance-wise evaluation}
\vspace{-.5em}

In the utterance-wise evaluation, each utterance is extracted by using ground-truth segmentation information. To obtain the segmentation for each far-field recording, 
the audio is first aligned with the corresponding close talking reference audio based on cross correlation. 
Then, each utterance is cut out from the far-field audio based on the ground truth utterance boundary information. 
Separation processing is performed on each utterance independently.
The separated signals are fed to the ASR system to calcualte WERs. 
For each input utterance, two transcriptions and WERs are generated. The one with a lower WER is picked.

\subsubsection{Continuous input evaluation}
\vspace{-.5em}

In the continuous input evaluation mode, separation and recognition are performed on an audio stream without splitting it into individual utterances. 
However, since online decoding support is relatively limited in many open source ASR frameworks, it is not straightforward to simply feed the entire audio of each mini session to the ASR and get reasonable results.  
To sidestep this, in this work, we opt to perform long-segment-wise decoding instead of truly continuous ASR, where we pre-segment each mini session into relatively long segments of 60 to 120 seconds in duration, based on the ground truth utterance boundaries. The segment boundaries are chosen such that they take place during silence. This results in segments each containing around 8 to 10 utterances.
ASR transcriptions must be generated for each long segment. 
Asclite\footnote{https://github.com/usnistgov/SCTK},
which can align multiple (two in this work) hypotheses against multiple reference transcriptions, 
is used to estimate (speaker agnostic) WERs. 

For both evaluation modes, Session0 in LibriCSS dataset can be used as the development set for hyper-parameter tuning.

\section{Experimental Results}
\vspace{-.5em}

\subsection{Speech separation method}
\vspace{-.3em}

\begin{figure}[t]
    \centering
    \includegraphics[scale=0.625]{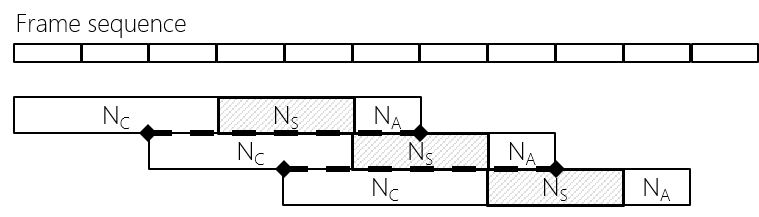}
    \vspace{-1em}
    \caption{Generating TF masks in a streaming fashion by using a sliding window. Output orders for each pair of neighboring chunks are aligned by using the frames that the two chunks share.}
    \label{fig: Blockproc}
\end{figure}

We report evaluation results of the speaker-independent CSS algorithm proposed in \cite{Yoshioka18b} to provide a reference point for calibrating the future results on this dataset. The method uses a bidirectional model to estimate three time-frequency (TF) masks: two for speech sources and one for noise, where the noise TF masks are needed for beamforming. 
To achieve this with streaming processing, which is required for the continuous input evaluation, 
a sliding window-based approach is employed as illustrated in Fig.~\ref{fig: Blockproc}.
The window comprises three subwindows, each representing a past, current, or future context. 
They consist of $N_{\past}$, $N_{\current}$, and $N_{\future}$ frames, respectively. 
At each operation point, the input features within the whole window are fed to the separation model to generate the TF masks. 
Then, only the TF masks within the current subwindow are used. 
The past and future subwindows help improve the mask estimation accuracy by providing left and right acoustic contexts to the separation model. 
The window is then shifted by $N_{\current}$ frames to process the next chunk of frames. 
Given the three TF masks,  
we output two speech signals with either spectral masking or mask-based adaptive minimum variance distortionless response (MVDR) beamforming, following \cite{PrincetonASRU2019}. 
For the utterance-wise evaluation, separation is performed with batch processing. 

For the continuous input evaluation, 
the CSS method described above generates two separated signals for each long segment (see Section 3.3.3) or mini session. 
In our ASR experiments, we further automatically segment each separated signal. 
A voice activity detector of https://github.com/wiseman/py-webrtcvad is used with mode ``0'', which ensures least aggressive non-speech filtering. 
The same procedure is consistently applied to the original center-microphone recordings, providing a baseline for the continuous evaluation.

Our separation model consists of three 1024-cell  BLSTM layers, followed by three parallel ReLU projection layers for mask estimation. The model 
is trained on 206 hours of artificially reverberated and mixed speech signals to minimize the Euclidean distance between reference and masked speech signals. The clean signals are randomly sampled from \texttt{train-clean-\{100,360\}}~\cite{panayotov2015librispeech}. 
For each mixture sample, room impulse responses are generated with the image method by assuming a random room with $T_{60} \in [0.15 \text{ secs}, 0.6 \text{ secs}]$.
Following \cite{habets2007generating}, 
multi-channel isotropic noise signals are generated  and added to the mixture signal at a random SNR from 0 to 10 dB. 

Four additional models are also built to examine the effect of the number and arrangement of microphones. One is based only on a single channel input. Two models are based on three-channel input: one uses microphones 1, 0, and 4 (i.e., a linear array) in Fig.~\ref{fig: data}; one uses microphones 1, 3, and 5 (i.e., a triangular array). 
The fourth model uses microphones 0, 1, 2, ,4, and 5.

\subsection{Utterance-wise evaluation}
\vspace{-.3em}

Table \ref{tab:res} shows the WERs for the utterance wise evaluation. 
We can see that even a small amount of overlap severely degraded the ASR performance. 
While the single-channel separation model improved the WERs when the overlap ratio was 10\% or higher, it degraded the WER when there was no speech overlap, which was not sufficiently considered in the prior studies as we discussed earlier. 
The seven-channel model clearly generated more accurate TF masks, outperforming the single-channel model in all conditions. 
However, TF masking did not improve the ASR accuracy for the non-overlap conditions even with the seven microphones. 
By contrast, the MVDR performance significantly surpassed that of the TF masking. This is in alignment with the observation in \cite{yoshioka2018multi}.
While the MVDR-based seven-channel system provided a substantial WER gain, the performance difference between the 0\% and 40\% overlap cases is still significant, calling for the development of more accurate separation algorithms.

\begin{table}[t]
\small
\centering
\caption{\small \%WERs for utterance-wise evaluation. 0S: 0\% overlap with short inter-utterance silence. 0L: 0\% overlap with long inter-utterance silence. Our ASR system yields WERs of 4.9\% and 5.1\% for anechoic versions of 0S and 0L utterances.}
\label{tab:res}
\vspace{.2em}
\begin{tabular}{lcccccc} \thickhline
\multirow{2}{*}{System} & \multicolumn{6}{c}{Overlap ratio in \%} \\
 &0S& 0L & 10 & 20 & 30& 40\\ \hline
No separation & 11.8 & 11.7 & 18.8 & 27.2 & 35.6& 43.3 \\
Mask (1ch) &  12.7 &  12.1 & 17.6 & 23.2 & 30.5& 35.6 \\
Mask (7ch) & 12.0 & 11.6 & 15.6& 20.2&25.6 &29.4  \\
MVDR (7ch) & 8.4 & 8.3 & 11.6 & 15.8 & 18.7 & 21.7 \\ \hline
\end{tabular}
\end{table}

\subsection{Continuous input evaluation}
\vspace{-.3em}

\begin{table}[t]
\small
\centering
\caption{\small \%WERs for seven-channel continuous input evaluation with different chunking configurations. The dash-separated three numbers of the first column are $N_{\past}$, $N_{\current}$, $N_{\future}$ values, respectively. Inherent latency is shown in parentheses.}
\label{tab:cssresults}
\vspace{.2em}
\begin{tabular}{lcccccc} \thickhline
\multirow{2}{*}{System} & \multicolumn{6}{c}{Overlap ratio in \%} \\
 &0S& 0L & 10 & 20 & 30& 40\\
\hline
 No separation & 15.4 & 11.5 & 21.7 & 27.0& 34.3& 40.5 \\ \hline
 1.2-0.8-0.4 (1.2 s) & 11.9 & 9.7 & 13.6& 15.0& 19.9& 21.9 \\
 1.6-0.8-0.0 (0.8 s) &  12.2 & 9.7& 14.7& 16.1& 20.5& 23.1 \\
 0.8-0.4-0.4 (0.8 s) & 11.5 & 9.5 & 13.4 & 15.8 & 19.7 & 21.2 \\
\hline
\end{tabular}
\end{table}

\begin{table}[t]
\small
\centering
\caption{\small \%WER impact of number and arrangement of microphones in continuous input case. $N_{\past}$, $N_{\current}$, and $N_{\future}$ are set at equivalents of 1.2 s, 0.8 s, and 0.4 s, respectively.}
\label{tab:cssresults_mics}
\vspace{.2em}
\begin{tabular}{lcccccc} \thickhline
\multirow{2}{*}{System} & \multicolumn{6}{c}{Overlap ratio in \%} \\
 &0S& 0L & 10 & 20 & 30& 40\\
\hline
 7ch & 11.9 & 9.7 & 13.6& 15.0& 19.9& 21.9 \\
 5ch & 12.8 & 10.5 & 15.3& 17.4& 22.8& 26.4 \\
 3ch (triangular) &  15.8 &  10.8 & 19.4 & 23.1 & 28.9 & 36.0 \\
 3ch (linear) & 15.1 & 9.8 & 17.7 & 20.6 & 30.1  & 29.6 \\
 1ch &  17.6 & 16.3  & 20.9  & 26.1  & 32.6 & 36.1 \\
\hline
\end{tabular}
\end{table}

Table \ref{tab:cssresults} shows the WERs for the continuous input evaluation with seven-channel input. 
It can be seen that the CSS algorithm improved the WERs for all conditions, including the cases where there was no speech overlap (i.e., 0S and 0L) thanks to the beamforming processing. 
The first system, denoted as 1.2-0.8-0.4, uses the same configuration as \cite{PrincetonASRU2019}, using the past, current, and future contexts of 1.2 s ($N_{\past}$), 0.8 s ($N_{\current}$), and 0.4 s ($N_{\future}$) , respectively.
This system has an inherent latency of 1.2 s (i.e., $N_{\current} + N_{\future}$). 
The inherent latency means the amount of delay caused by the system configuration, which does not include the processing delay resulting from actual computation for network evaluation, MVDR computation, and so on.  
In the last two rows of Table~\ref{tab:cssresults}, 
two different configurations are examined, both of which reduces the inherent latency to 0.8 s. One way (1.6-0.8-0.0) is to avoid look-ahead by setting $N_{\future}$ at 0. The chunk size i.e., $N_{\past} + N_{\current} + N_{\future}$, is kept constant. 
The other way (0.8-0.4-0.4) is to reduce the chunk size while keeping the look-ahead size constant. 
The results clearly show the benefit of taking account of the future acoustic context (i.e., keeping $N_{\future}$ at the equivalent of 0.4 s). 

We can also see the detrimental effect of turn-taking without much silence. In the baseline system (i.e., without separation processing), the WER was increased by 33.9\% relative just by reducing the inter-utterance gap from around 3 s (0L) to 0.5 s or shorter (0S) even though their WERs were almost the same when correct segmentations were provided (see Table~\ref{tab:res}). The CSS processing also mitigated this degradation by routing the adjacent utterances to different output channels, which reduced the relative WER increase to 22.7\%.

Table \ref{tab:cssresults_mics} lists the WERs for different microphone setups. 
The results clearly show the significant impact that the number of microphones has on the performance. No meaningful improvement was observed for the non-overlapping setting when three microphones were used. 
As with the utterance-wise evaluation results, 
the single-channel system even degraded the WERs for this setting while it still slightly provided gains for the other settings.

\section{Conclusion}
\vspace{-.5em}

This paper described a dataset and protocols for evaluating continuous speech separation algorithms. The dataset is called LibriCSS and consists of multi-channel recordings of LibriSpeech utterances concatenated and replayed in a meeting room. 
By using a PIT-based speaker-independent CSS method, several aspects of CSS are investigated on this dataset. 

Our experimental results shed light on the areas that need further improvement. Firstly, even with seven-channel MVDR, the performance degradation caused by speech overlap is not trivial. Also, in the single- and three-microphone cases, the separation processing does not improve or sometimes degrades the WER when only one speaker is active. 
We hope that this dataset and the associated evaluation pipeline facilitate speech separation research while helping reduce the gap between the research state and what is required in real conversational processing applications. 


\newpage

\bibliographystyle{IEEEtran}
\bibliography{bibtex}

\begin{thebibliography}{10}
\providecommand{\url}[1]{#1}
\csname url@samestyle\endcsname
\providecommand{\newblock}{\relax}
\providecommand{\bibinfo}[2]{#2}
\providecommand{\BIBentrySTDinterwordspacing}{\spaceskip=0pt\relax}
\providecommand{\BIBentryALTinterwordstretchfactor}{4}
\providecommand{\BIBentryALTinterwordspacing}{\spaceskip=\fontdimen2\font plus
\BIBentryALTinterwordstretchfactor\fontdimen3\font minus
  \fontdimen4\font\relax}
\providecommand{\BIBforeignlanguage}[2]{{%
\expandafter\ifx\csname l@#1\endcsname\relax
\typeout{** WARNING: IEEEtran.bst: No hyphenation pattern has been}%
\typeout{** loaded for the language `#1'. Using the pattern for}%
\typeout{** the default language instead.}%
\else
\language=\csname l@#1\endcsname
\fi
#2}}
\providecommand{\BIBdecl}{\relax}
\BIBdecl

\bibitem{hershey2016deep}
J.~R. Hershey, Z.~Chen \emph{et~al.}, ``Deep clustering: discriminative
  embeddings for segmentation and separation,'' in \emph{Proc. ICASSP 2016},
  2016, pp. 31--35.

\bibitem{yu2017permutation}
D.~Yu, M.~Kolb{\ae}k \emph{et~al.}, ``Permutation invariant training of deep
  models for speaker-independent multi-talker speech separation,'' in
  \emph{Proc. ICASSP 2017}, 2017, pp. 241--245.

\bibitem{wang2018alternative}
Z.-Q. Wang, J.~Le~Roux \emph{et~al.}, ``Alternative objective functions for
  deep clustering,'' in \emph{Proc. ICASSP 2018}, 2018, pp. 686--690.

\bibitem{xu2019optimization}
C.~Xu, W.~Rao \emph{et~al.}, ``Optimization of speaker extraction neural
  network with magnitude and temporal spectrum approximation loss,'' in
  \emph{Proc. ICASSP 2019}.\hskip 1em plus 0.5em minus 0.4em\relax IEEE, 2019,
  pp. 6990--6994.

\bibitem{chen2017deep}
Z.~Chen, Y.~Luo \emph{et~al.}, ``Deep attractor network for single-microphone
  speaker separation,'' in \emph{Proc. ICASSP 2017}.\hskip 1em plus 0.5em minus
  0.4em\relax IEEE, 2017, pp. 246--250.

\bibitem{chen2018sequence}
Z.~Chen and J.~Droppo, ``Sequence modeling in unsupervised single-channel
  overlapped speech recognition,'' in \emph{Proc. ICASSP 2018}, 2018, pp.
  4809--4813.

\bibitem{li2019listening}
Z.-X. Li, Y.~Song \emph{et~al.}, ``Listening and grouping: an online
  autoregressive approach for monaural speech separation,'' \emph{IEEE/ACM
  TASLP}, vol.~27, no.~4, pp. 692--703, 2019.

\bibitem{wang2019pitch}
K.~Wang, F.~Soong \emph{et~al.}, ``A pitch-aware approach to single-channel
  speech separation,'' in \emph{Proc. ICASSP 2019}, 2019, pp. 296--300.

\bibitem{ephrat2018looking}
A.~Ephrat, I.~Mosseri \emph{et~al.}, ``Looking to listen at the cocktail party:
  A speaker-independent audio-visual model for speech separation,''
  \emph{arXiv:1804.03619}, 2018.

\bibitem{zhao2018sound}
H.~Zhao, C.~Gan, A.~Rouditchenko \emph{et~al.}, ``The sound of pixels,''
  \emph{arXiv:1804.03160}, 2018.

\bibitem{zmolikova2017speaker}
K.~Zmolikova, M.~Delcroix \emph{et~al.}, ``Speaker-aware neural network based
  beamformer for speaker extraction in speech mixtures,'' in
  \emph{Interspeech}, 2017, pp. 2655--2659.

\bibitem{wang2018voicefilter}
Q.~Wang, H.~Muckenhirn \emph{et~al.}, ``{VoiceFilter}: targeted voice
  separation by speaker-conditioned spectrogram masking,''
  \emph{arXiv:1810.04826}, 2018.

\bibitem{xiao2019single}
X.~Xiao, Z.~Chen \emph{et~al.}, ``Single-channel speech extraction using
  speaker inventory and attention network,'' in \emph{Proc. ICASSP 2019}, 2019,
  pp. 86--90.

\bibitem{chen2018location}
Z.~Chen, X.~Xiao \emph{et~al.}, ``Multi-channel multi-speaker overlapped speech
  recognition with location guided speech extraction network,'' in \emph{Proc.
  SLT 2018}, 2018, pp. 558--565.

\bibitem{perotin2018multichannel}
L.~Perotin, R.~Serizel \emph{et~al.}, ``Multichannel speech separation with
  recurrent neural networks from high-order ambisonics recordings,'' in
  \emph{Proc. ICASSP 2018}, 2018, pp. 36--40.

\bibitem{zhao2018two}
Y.~Zhao, Z.-Q. Wang \emph{et~al.}, ``Two-stage deep learning for
  noisy-reverberant speech enhancement,'' \emph{IEEE/ACM TASLP}, vol.~27,
  no.~1, pp. 53--62, 2018.

\bibitem{wang2018multi}
Z.-Q. Wang, J.~Le~Roux \emph{et~al.}, ``Multi-channel deep clustering:
  discriminative spectral and spatial embeddings for speaker-independent speech
  separation,'' pp. 1--5, 2018.

\bibitem{yoshioka2018multi}
T.~Yoshioka, H.~Erdogan \emph{et~al.}, ``Multi-microphone neural speech
  separation for far-field multi-talker speech recognition,'' in \emph{Proc.
  ICASSP 2018}, 2018, pp. 5739--5743.

\bibitem{luo2018tasnet}
Y.~Luo and N.~Mesgarani, ``{TasNet}: time-domain audio separation network for
  real-time, single-channel speech separation,'' in \emph{Proc. ICASSP 2018},
  2018, pp. 696--700.

\bibitem{luo2019conv}
Y.~Luo and N.~{Mesgarani}, ``{Conv-TasNet}: surpassing ideal time--frequency
  magnitude masking for speech separation,'' \emph{IEEE/ACM TASLP}, vol.~27,
  no.~8, pp. 1256--1266, 2019.

\bibitem{shi2019furcanext}
Z.~Shi, H.~Lin \emph{et~al.}, ``{FurcaNeXt}: End-to-end monaural speech
  separation with dynamic gated dilated temporal convolutional networks,''
  \emph{arXiv:1902.04891}, 2019.

\bibitem{fevotte2005bss_eval}
C.~F{\'e}votte, R.~Gribonval \emph{et~al.}, ``{BSS\_EVAL} toolbox user
  guide---revision 2.0,'' Tech. Rep. inria-00564760, 2005.

\bibitem{le2019sdr}
J.~Le~Roux, S.~Wisdom \emph{et~al.}, ``{SDR}---half-baked or well done?'' in
  \emph{Proc. ICASSP 2019}, 2019, pp. 626--630.

\bibitem{luo2019dualpath}
Y.~Luo, Z.~Chen \emph{et~al.}, ``Dual-path {RNN}: efficient long sequence
  modeling for time-domain single-channel speech separation,'' 2019.

\bibitem{Cetin06}
O.~\c{C}etin and E.~Shriberg, ``Analysis of overlaps in meetings by dialog
  factors, hot spots, speakers, and collection site: insights for automatic
  speech recognition,'' in \emph{Proc. Interspeech}, 2006, pp. 293--296.

\bibitem{HoriMeeting}
T.~{Hori}, S.~{Araki} \emph{et~al.}, ``Low-latency real-time meeting
  recognition and understanding using distant microphones and omni-directional
  camera,'' \emph{IEEE TASLP}, vol.~20, no.~2, pp. 499--513, Feb 2012.

\bibitem{NTTAllNeural}
T.~{von Neumann}, K.~{Kinoshita} \emph{et~al.}, ``All-neural online source
  separation, counting, and diarization for meeting analysis,'' in \emph{Proc.
  ICASSP}, 2019, pp. 91--95.

\bibitem{Kanda2019}
N.~{Kanda}, S.~{Horiguchi} \emph{et~al.}, ``Simultaneous speech recognition and
  speaker diarization for monaural dialogue recordings with target-speaker
  acoustic models,'' \emph{arXiv:1909.08103}, 2019.

\bibitem{PrincetonASRU2019}
T.~Yoshioka, I.~Abramovski \emph{et~al.}, ``Advances in online audio-visual
  meeting transcription,'' in \emph{Proc. ASRU}, 2019, pp. 276--283.

\bibitem{wichern2019wham}
G.~Wichern, J.~Antognini \emph{et~al.}, ``Wham!: extending speech separation to
  noisy environments,'' \emph{arXiv:1907.01160}, 2019.

\bibitem{bahmaninezhad2019comprehensive}
F.~Bahmaninezhad, J.~Wu \emph{et~al.}, ``A comprehensive study of speech
  separation: spectrogram vs waveform separation,'' \emph{arXiv:1905.07497},
  2019.

\bibitem{barker2018fifth}
J.~Barker, S.~Watanabe \emph{et~al.}, ``The fifth `chime' speech separation and
  recognition challenge: dataset, task and baselines,''
  \emph{arXiv:1803.10609}, 2018.

\bibitem{van2019dipco}
M.~Van~Segbroeck, A.~Zaid \emph{et~al.}, ``{DiPCo}--dinner party corpus,''
  \emph{arXiv:1909.13447}, 2019.

\bibitem{ICSICorpus}
A.~{Janin}, D.~{Baron} \emph{et~al.}, ``The {ICSI} meeting corpus,'' in
  \emph{Proc. ICASSP 2003}, 2003, pp. 364--367.

\bibitem{AMICorpus}
J.~Carletta, S.~Ashby \emph{et~al.}, ``The {AMI} meeting corpus: a
  pre-announcement,'' in \emph{Proc. ICMI-MLMI 2006}, 2006, pp. 28--39.

\bibitem{povey2011kaldi}
D.~Povey, A.~Ghoshal \emph{et~al.}, ``The kaldi speech recognition toolkit,''
  in \emph{Proc. ASRU}.\hskip 1em plus 0.5em minus 0.4em\relax IEEE, 2011.

\bibitem{lu2019pykaldi2}
L.~Lu, X.~Xiao \emph{et~al.}, ``{PyKaldi2}: yet another speech toolkit based on
  {Kaldi} and {PyTorch},'' \emph{arXiv:1907.05955}, 2019.

\bibitem{Yoshioka18b}
T.~Yoshioka, H.~Erdogan \emph{et~al.}, ``Recognizing overlapped speech in
  meetings: a multichannel separation approach using neural networks,'' in
  \emph{Proc. Interspeech}, 2018, pp. 3038--3042.

\bibitem{panayotov2015librispeech}
V.~Panayotov, G.~Chen \emph{et~al.}, ``Librispeech: an {ASR} corpus based on
  public domain audio books,'' in \emph{Proc. ICASSP 2015}, 2015, pp.
  5206--5210.

\bibitem{habets2007generating}
E.~A. Habets and S.~Gannot, ``Generating sensor signals in isotropic noise
  fields,'' \emph{JASA}, vol. 122, no.~6, pp. 3464--3470, 2007.

\end{thebibliography}

\end{document}